\documentclass[twocolumn,showpacs,preprintnumbers,amsmath,amssymb,pre,superscriptaddress]{revtex4-1} 
\usepackage{graphicx} 
\usepackage{dcolumn}
\usepackage{bm} 
\usepackage{subfigure} 
\usepackage{epstopdf}
\usepackage[usenames]{color} 
\usepackage{float} 
\usepackage{CJK}

\begin{document}
\begin{CJK}{UTF8}{gbsn}

\title{Firing dynamics of an autaptic neuron}


\author{Hengtong Wang}
\affiliation{School of Physics and Information Technology, Shaanxi Normal University, Xi'an 710119, China}

\author{Yong Chen}
\email{ychen@buaa.edu.cn}
\affiliation{Center of soft matter physics and its application, Beihang University, Beijing 100191, China\\
School of Physics and Nuclear Energy Engineering, Beihang University, Beijing 100191, China}

\begin{abstract}
Autapses are synapses that connect a neuron to itself in the nervous system. Previously, both experimental and theoretical studies have demonstrated that autaptic connections in the nervous system have a significant physiological function. Autapses in nature provide self-delayed feedback, thus introducing an additional time scale to neuronal activities and causing many dynamic behaviors in neurons. Recently, theoretical studies have revealed that an autapse provides a control option for adjusting the response of a neuron: e.g., an autaptic connection can cause the electrical activities of the Hindmarsh-Rose neuron to switch between quiescent, periodic, and chaotic firing patterns; an autapse can enhance or suppress the mode-locking status of a neuron injected with sinusoidal current; and the firing frequency and interspike interval distributions of the response spike train can also be modified by the autapse. In this paper, we review recent studies that showed how an autapse affects the response of a single neuron.  


\end{abstract}
\maketitle

\section{Introduction}
The nervous system is a complex network of neurons, synapses, and other specialized cells, and through this system, neurons receive and transmit information between different parts of the body~\cite{kandel}. The responses of neurons and the function of synapses have received considerable attention because of their status as building blocks of the nervous system~\cite{Bartos,Bennett,Connors}. Nearly a century ago, neuroscientists found a special synaptic structure, the autapse, which is a synapse between different parts of the same neuron~\cite{vander,Bekkers1998,yamaguichi}. Because of its odd structure, the self-synaptic connection has remained poorly understood. However, many recent studies found that autapses are much more widespread in the nervous system than previously thought. Autapses have been reported in various brain areas, such as the neocortex, cerebellum, hippocampus, striatum, and substantia nigra~\cite{lubke,Bekkers1998,flight,branco,Kimura,tamas}. Interestingly, about 80\% of cortical pyramidal neurons have autaptic connections~\cite{lubke}. Many studies have revealed that autapses are not merely curiosities, but play an authentic physiological role in the nervous system. Autapses can maintain persistent activity in the nervous system by mediating positive feedback~\cite{saada}.

Recent experiments also demonstrated that autapses are very important for the processing function of the brain~\cite{ikeda}. Bacci et al. recorded the activity of fast spiking interneurons in acute brain slices of juvenile rats and found that autaptic transmission increased the spike-timing precision~\cite{bacci}. A somatic spike can evoke excitatory postsynaptic currents with various amplitudes through an autapse~\cite{Kimura}. However, the function of autapses and their contribution to information processing are still unclear~\cite{ikeda,bekkers09}. Thus, understanding the effect of autaptic activities on the responses of a neuron is a fundamental step to elucidating the process of information transfer in the nervous system~\cite{salinas1,ychen2008,chenyl,quiroga}. 

In fact, natural autapses are self-feedback connections in the nervous system and may allow a unique type of self-control, similar to the feedback in other systems~\cite{ikeda}. Actually, feedback creates a circuit, or loop, that connects a system to itself and commonly occurs in many systems, e.g., gene regulatory networks~\cite{davidson,zhanghui}, population dynamics~\cite{cabrera}, and climate systems~\cite{bonan}. Feedback is also used extensively to control the state of a nonlinear system, such as to stabilize periodic orbits or to control coherence resonance~\cite{post,sethia}. Feedback is also very commonly used in the design of electric circuit elements~\cite{waik}. In these systems, information about the past or the present influences the same phenomena in the present or future, respectively. Moreover, feedback has a marked effect on the dynamics of nonlinear systems. The frequency at the onset of the oscillation of a nonlinear system can be modified by the feedback loop~\cite{Gaudreault}. Furthermore, self-delayed feedback with a fixed delay time can suppress chaos and also control steady states~\cite{Ahlborn,Balanov}. In a simulation study, Popovych et al. showed that time-delayed feedback has the ability to desynchronize groups of model neurons~\cite{popovych}.

A neuron with an autaptic connection can provide a distinctive physiological self-feedback model with many dynamic properties. Rusin et al. performed a time-delayed feedback stimulation of a group of cultured neurons and experimentally demonstrated that time-delayed feedback could cause synchronization of the action potentials of a group of neurons~\cite{rusin}. Prager et al. found that time-delayed feedback can facilitate the noise-induced oscillation of a neuronal system~\cite{prager}. Autaptic delayed feedback was found to modify the bursting in the distribution of interspike intervals of a stochastic Hodgkin-Huxley (HH) neuron~\cite{liyy}. Such autaptic delayed feedback can also reduce the spontaneous spiking activity of a stochastic neuron at the characteristic frequencies. The effect of an autapse on the spike rate of a single neuronal system shows dependence on the duration of the autapse activity~\cite{hashemi}. Modulation of autaptic delay feedback can cause the dynamic behaviors of a Hindmarsh-Rose (HR) neuron to switch between quiescent, periodic, and chaotic firing patterns~\cite{wht2014a}. 

In this article, we review the current stat of knowledge of the effect of an autapse on single neurons. 

\section{Autaptic connection}

\begin{figure}
\begin{center}
\includegraphics[width=0.5\textwidth]{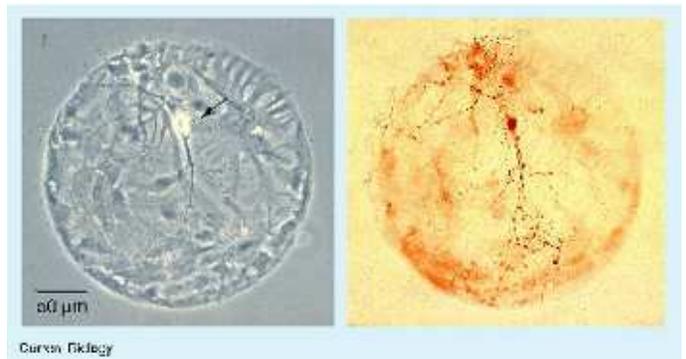}
\end{center}
\caption{A single rat hippocampal neuron (arrowed, left) grown in culture on a ­‘microdot’ of glia (flat gray cells), showing abundant autapses labeled with an antibody (small dark spots, right)~\cite{ikeda}.} 
\label{autapse2}
\end{figure}

Scientists have long investigated the functions and properties of neurons and synapses. Neurons vary widely with different sizes and forms, and neuron responses are also involved in many dynamic behaviors. A synapse is a structure in the nervous system that permits a neuron (or nerve cell) to transmit an electrical or chemical signal to another cell (neural or otherwise)~\cite{schacter}. The self-synapse, or the autapse, was first described as a synapse between the axon of a pyramidal cell and its own dendrites by Van der Loos and Glaser in 1972~\cite{vander}. Before the report by Van der loos et al., other terms, such as 'self-excitation' and 'self-sensing', were used to describe the self-synaptic structure. For a long time, autapses in the nervous system seemed like anatomical curiosities with questionable functional significance. However, recent experiments began to reveal how autapses might play an important role in brain function. Some reports also suggested that autapses are connected to some neural diseases~\cite{jang}. Neurobiology experiments have always focused on excitatory (glutamate-releasing) and inhibitory (GABA-releasing) autapses. To date, autaptic structures have been found in almost all parts of the nervous system~\cite{Bekkers1998}. Fig.~\ref{autapse2} shows photos of the many autapses of a single rat hippocampal neuron. 

\begin{figure}
\begin{center}
\includegraphics[width=0.45\textwidth]{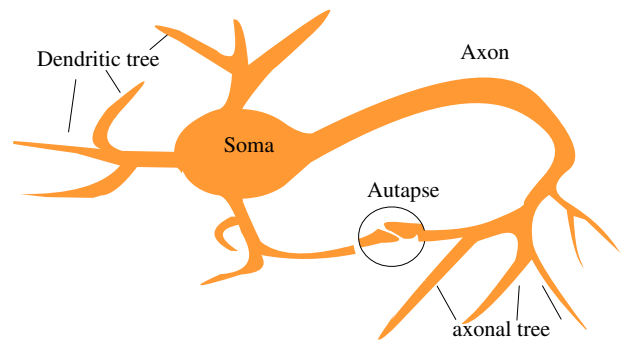}
\end{center}
\caption{Schematic drawing of a neuron with autaptic feedback. The structure in the circle represents an autaptic connection between dendritic and axonal terminals. } 
\label{autapse3}
\end{figure}

In nature, autaptic connections enable self-feedback of a neuron. Thus, in almost all current autapse studies, the model of a neuron with an autaptic connection contains a single compartment that exhibits a self-delayed feedback mechanism, as shown in Fig.~\ref{autapse3}. The delay time represents the elapsed time associated with the axonal propagation prior to the reintroduction of the signal into the neuron. The delay time is also believed to be one of the important properties of autaptic connections. Mathematical models of an autapse in previous studies of the effect of self-synapses on neurons fall into two main categories: one has linear self-coupling and describes a gap junction, and the other has nonlinear self-coupling and describes a chemical synapse, which may be an excitatory or inhibitory synapse. The specific mathematical form of the autapse depends on the chosen neuronal model. 

The autaptic connection of a simplified neuron model (such as the HR neuron, FitzHugh-Nagumo neuron, Izhikevich neuron and others) can often be described by one of the following formulas~\cite{hashemi,wht2014a,majun1}:

\begin{itemize}
\item Linear coupling gives as electrical diffusive-type:
\begin{equation}
I_{aut}=g_{aut}[u(t-\tau)-u(t)]. 
\label{eq1}
\end{equation}
$g_{aut}$ is the autaptic conductivity. $\tau$ is the delay time. This form of autaptic current is proportional to the difference between the membrane potential at $t$ and that at an earlier time point $t-\tau$.

\item The chemical synapse function is modeled using the so-called fast threshold modulation (FTM) scheme~\cite{Belykh,buric}:
\begin{eqnarray}
I_{aut}(t)&=&-g_{aut}(u(t)-V_{syn})S(t-\tau), \\
S(t-\tau)&=&1/\{1+\exp[-k(u(t-\tau)-\theta)]\},
\end{eqnarray}
where $g_{aut}$ is the autaptic intensity, and $V_{syn}$ is the synaptic reversal potential. $V_{syn}$=2 and $V_{syn}$=-2 correspond to excitatory and inhibitory autapses, respectively.
\end{itemize}

For a conductance-based neuron model (such as the HH neuron, Morris-Lecar neuron, Connor-Stevens neuron and others), the mathematic model of the autapse is often given as follows~\cite{liyy,hashemi,wht_jtb,wht_chaos}
\begin{itemize}
\item Linear coupling is also always set as electrical diffusive-type and has the same form as Eq.~\ref{eq1}.
\item Conductance-based chemical autapses can be described using a bi-exponential function, which has been reported to fit well with experiment results~\cite{william},
 \begin{equation}
 I_{aut}(t)=G(t-t_{delay})(V-E_{syn}),
\label{chemsyn}
 \end{equation}
where $G(t-t_{delay})$ is the autaptic conductance, and $E_{syn}$ is the autaptic reversal potential. In this equation, we used $E_{syn}=0\;\mathrm{mV}$ for excitatory neurons and -80 $\mathrm{mV}$ for inhibitory neurons. The autaptic conductance is defined as
\begin{eqnarray}
G(t)&=&g_{aut}\sigma(t-t_{fire})\\
\sigma(t)&=&\frac{\exp(-t/t_d)-\exp(-t/t_r)}{t_d-t_r}
\end{eqnarray}
with presynaptic spikes occurring at $t_{fire}$. The parameters $t_d$ and $t_r$ represent the decay and rise times of the function, respectively, and these parameters determine the duration of the response. The autaptic rise and decay times were set to $t_r=0.1\;\mathrm{ms}$ and $t_d=3\;\mathrm{ms}$, respectively. 
\end{itemize}

In the following sections, we will review the two main forms of an autapse. Studies have shown that autapses that exhibit delayed feedback provide a control option for adjusting neuron responses.

\section{Firing pattern transition in a bursting neuron}

Experimental observations have indicated that action potentials can occur with different firing patterns, and the primary firing patterns are spiking and bursting~\cite{cocatre,mainen,izikevich2000,kepecs2002,guhuaguang,yuht}. Bursting is an extremely diverse general phenomenon in firing patterns exhibited by neurons in the central nervous system and spinal cord~\cite{wagenaar, wangj}. Previous studies on a bursting neuron with an autapse showed the novel dynamics and the transition of firing patterns induced by an autapse~\cite{wht2014a}.

The simplified model of Hindmarsh and Rose has turned out to capture the features of experimentally measured electrical data quite accurately, particularly for studies of the spiking-bursting behavior of neuron membrane potentials. 

Without an autapse, HR neurons exhibit many dynamic behaviors, including quiescent, regular spiking, periodic, and chaotic bursting firing patterns (see Fig.~1 in Ref.~\cite{wht2014a} and Ref.~\cite{tang}).

\begin{figure}
\begin{center}
\includegraphics[width=0.5\textwidth]{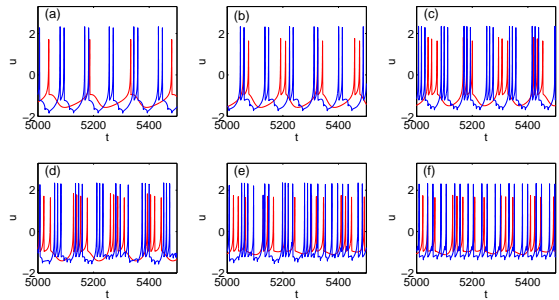}
\end{center}
\caption{Time courses of the membrane potentials in response to different $I_{ext}$ with $g_{aut}=0.5$~\cite{wht2014a}. The blue curves represent the time course of the HR neuron with an autapse, and the red curves indicate that without autapse.}
\label{fig3_hr_e1}
\end{figure}

The presence of an autapse completely changes the firing patterns of the original HR neuron. The firing pattern can be adjusted from a periodic or chaotic pattern to another periodic pattern or to a chaotic bursting pattern as the autaptic parameters change, independently of the original firing pattern. Fig.~\ref{fig3_hr_e1} shows the time courses of the membrane potentials of an HR neuron with an electrical autapse as an example. Importantly, the maximum action potential is increased by the electrical autapse. 

\begin{figure}
\begin{center}
\includegraphics[width=0.45\textwidth]{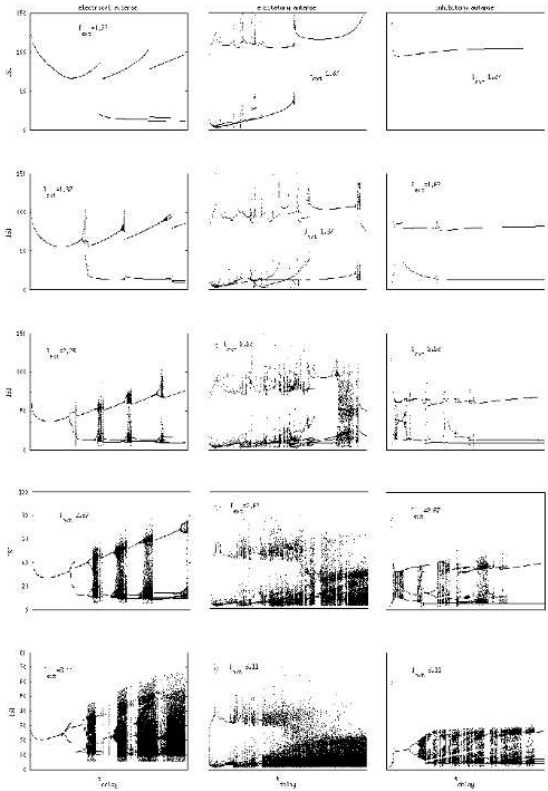}
\end{center}
\caption{Bifurcation diagram of the interspike interval of an HR neuron with an autapse versus the autaptic delay time~\cite{wht2014a}. From left to right, the panels show the results of the electrical autapse, excitatory chemical autapse, and inhibitory chemical autapse.}
\label{hr_e2}
\end{figure}

The ISI plot bifurcation diagrams of the HR neuron clearly show the transition between periodic and aperiodic firing. For an electrical autapse, the periodic state transits to the chaotic state exhibiting an alternating behavior as the time delay increased. With higher autaptic intensity, this alternating behavior is more noticeable and occurs more frequently. For shorter delay times, the firing pattern of the HR neuron showed periodic spiking independent of the external DC input. The neuron with an excitatory chemical autapse exhibits chaotic firing patterns in a larger area of $g_{aut}$-$\tau$ space than that of the neuron with an electrical autapse. For strong external stimuli, the chaotic region of the combinational parameters in the $g_{aut}$-$\tau$ space is enlarged. The excitatory autapse plays a positive role in generating and enhancing chaos as a whole. As neurons with an inhibitory autaptic connection, the chaotic spiking of the HR neuron can be decreased and suppressed. With the proper inhibitory autaptic parameters, the HR neuron could be driven to a resting state. Fig.~\ref{hr_e2} shows the bifurcation of the ISIs of an HR neuron for the three types of autapses. 

\begin{figure}
\begin{center}
\includegraphics[width=0.5\textwidth]{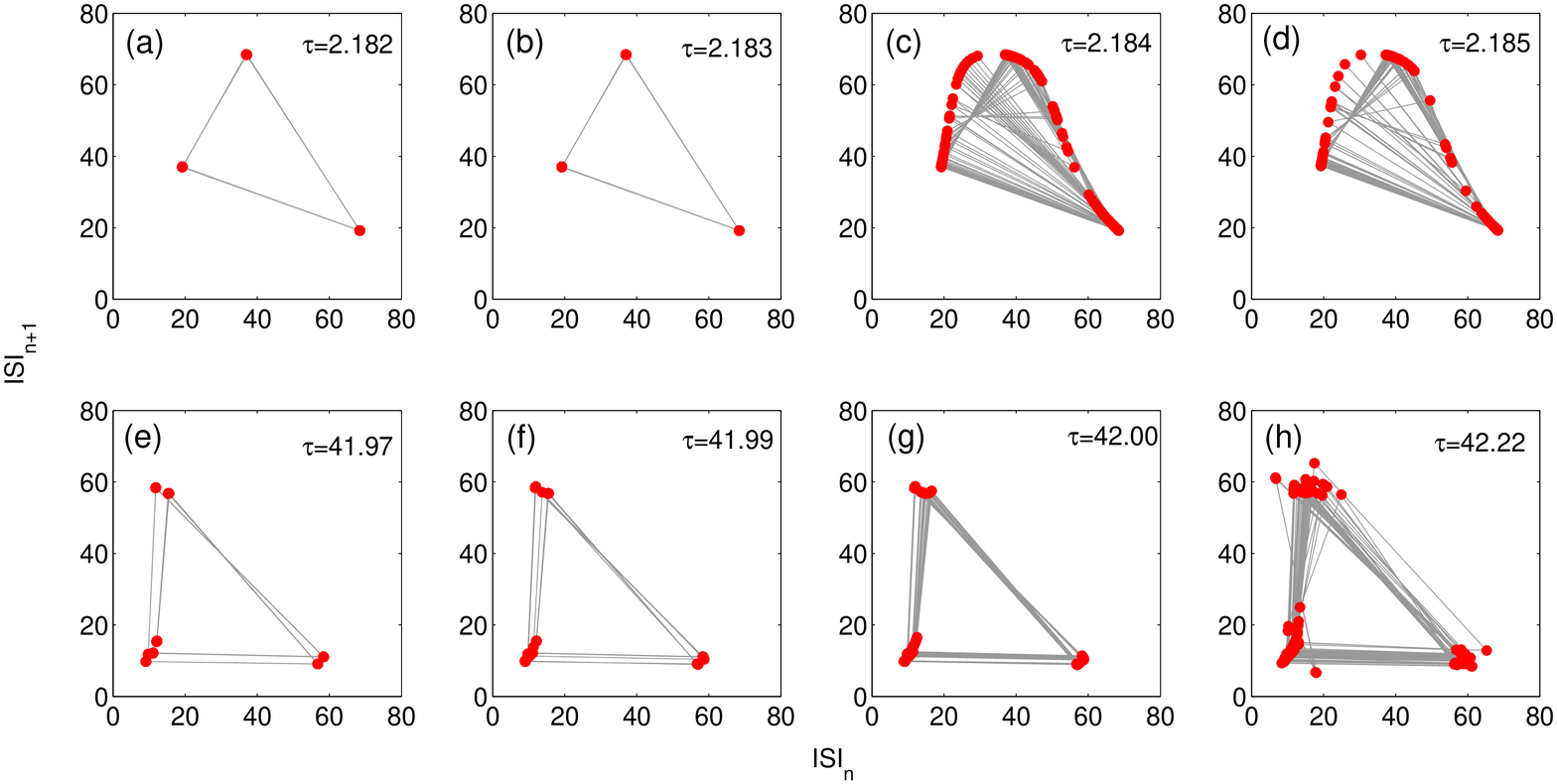}
\caption{An example of the two ways of transition (the discontinuous transition [upper] and the continuous transition [bottom]) into chaos from periodic firing~\cite{wht2014a}.
External stimuli is set as $I_{ext}$=2.67.}
\label{cnsns_returnmap1}
\end{center}
\end{figure} 

Without an autapse, the firing pattern of an HR neuron switches from silent to periodic bursting and then to chaotic firing with period-doubling bifurcation as the DC current increased~\cite{wangxj,gnzalez,innocenti,innocenti2009}. For neurons with an autapse, the firing pattern can switch into any other firing pattern, regardless of the original pattern. As a whole, there are two main approaches to transition from a periodic to a chaotic firing pattern: discontinuous and continuous. The interspike interval return map is a useful approach to characterizing the transition to chaos. Fig.~\ref{cnsns_returnmap1} shows an example of the two approaches for the transition of a neuron to chaos. In the discontinuous transition, the neuron has a periodic bursting firing pattern because the delay time is short. When the delay time increases, the system suddenly enters a chaotic firing state. In the continuous transition, the neuron first displays a periodic firing pattern. As the autaptic parameters change, the number of spikes in the single burst as well as the period of the periodic bursting increase almost continuously, and the neuron transitions into the chaotic state.

\section{Mode-locking behavior of a regular spiking neuron}

When excited by a periodic stimulus, neurons respond with various mode-locking firing patterns and quasi-periodic states~\cite{leesg,chey,wanght,fellous,wht_jtb}. When an autapse is present, the mode-locking firing of a neuron can be switched to another state, depending on the chosen autapse parameters. An autapse provides self-feedback and contributes an additional time scale to the dynamic neuronal system. Thus, the neuron-autapse system may exhibit very complex dynamics, due to the interplay between autaptic delayed feedback, an external periodic stimulus, and the intrinsic activity of the neuron.

In previous studies, the model neuron-autapse system often contained an HH neuron and an autapse. The HH model is a conductance-based model that describes how action potentials in neurons are initiated and propagated~\cite{hodgkin}. In this section, we review the mode-locking firing pattern of an HH neuron with an electrical autapse that has the same mathematical form as Eq.~(\ref{eq1}). 

\begin{figure}
\begin{center}
\includegraphics[width=0.23\textwidth]{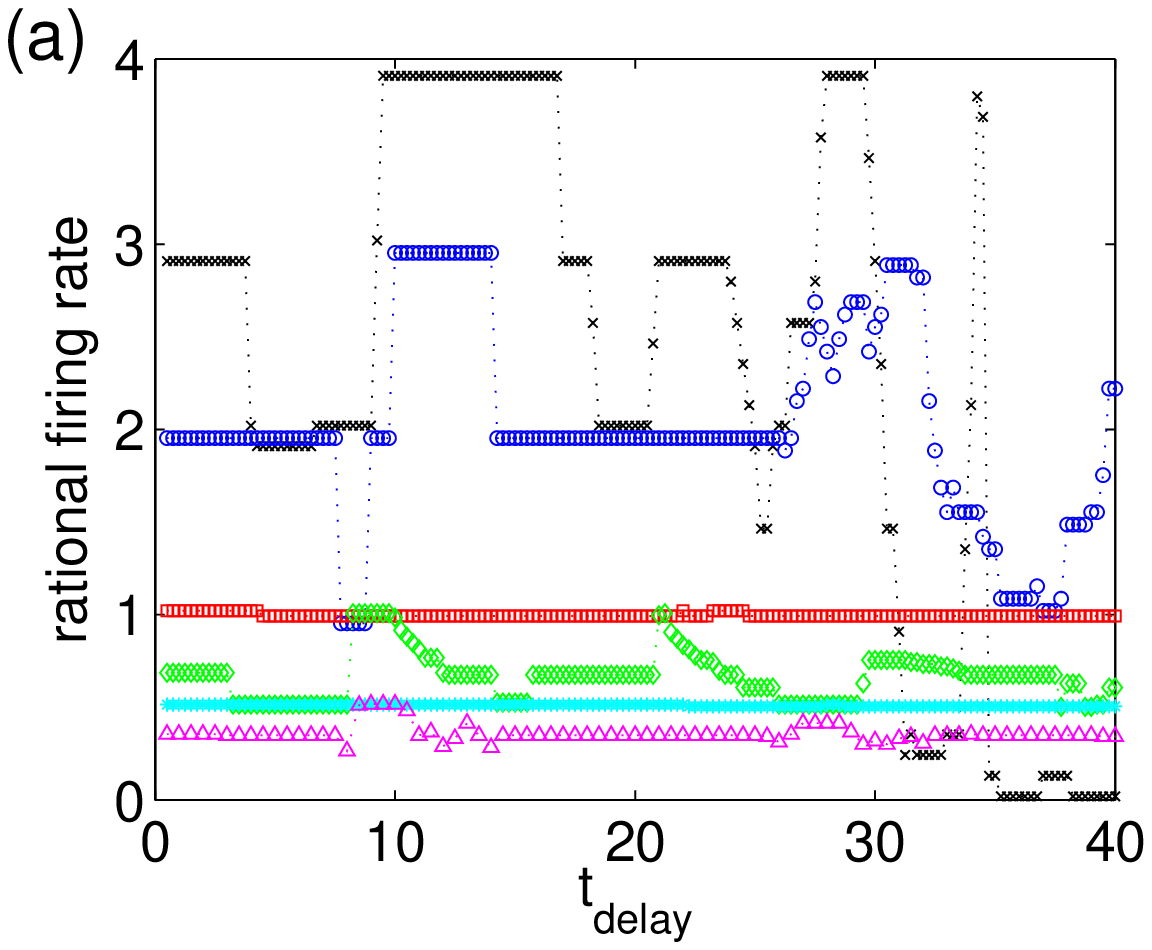}
\includegraphics[width=0.23\textwidth]{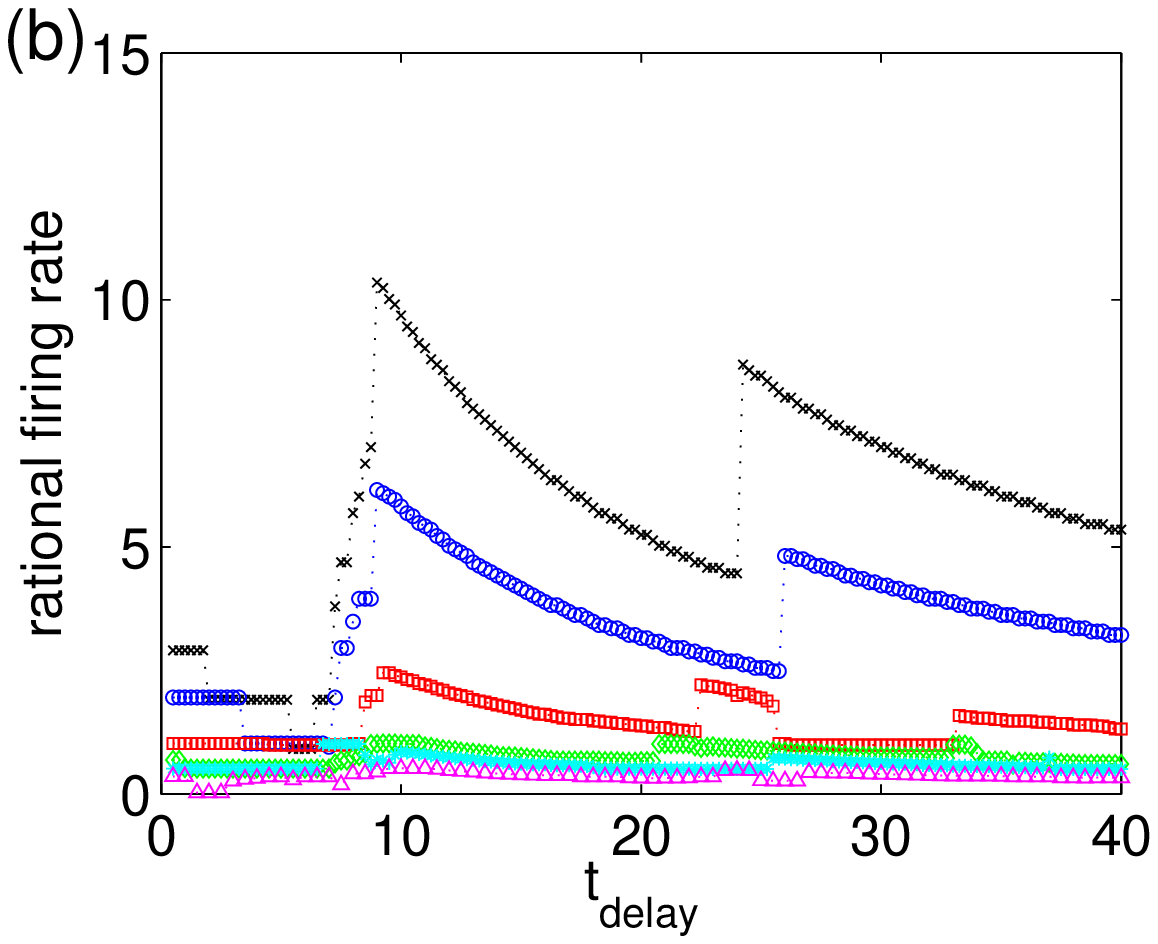}

\includegraphics[width=0.23\textwidth]{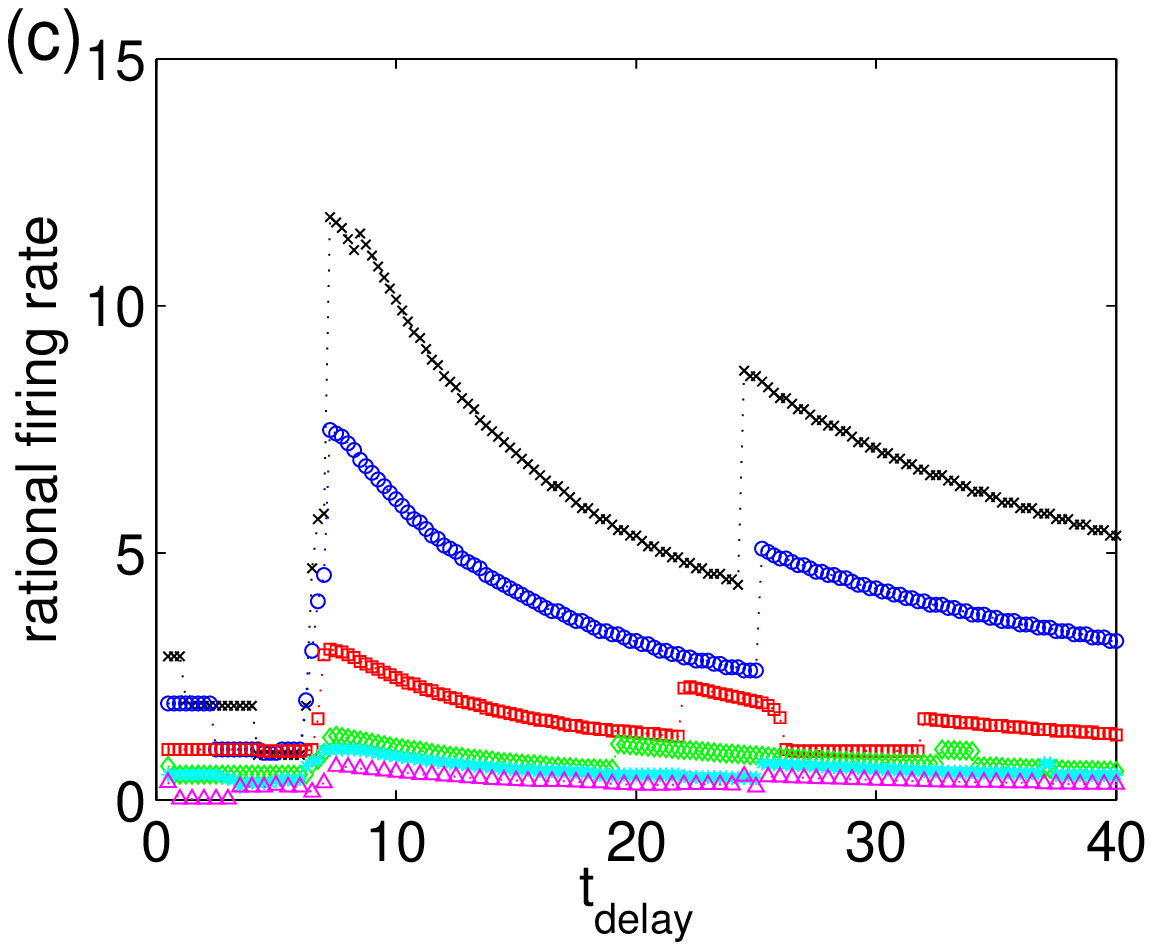}
\includegraphics[width=0.23\textwidth]{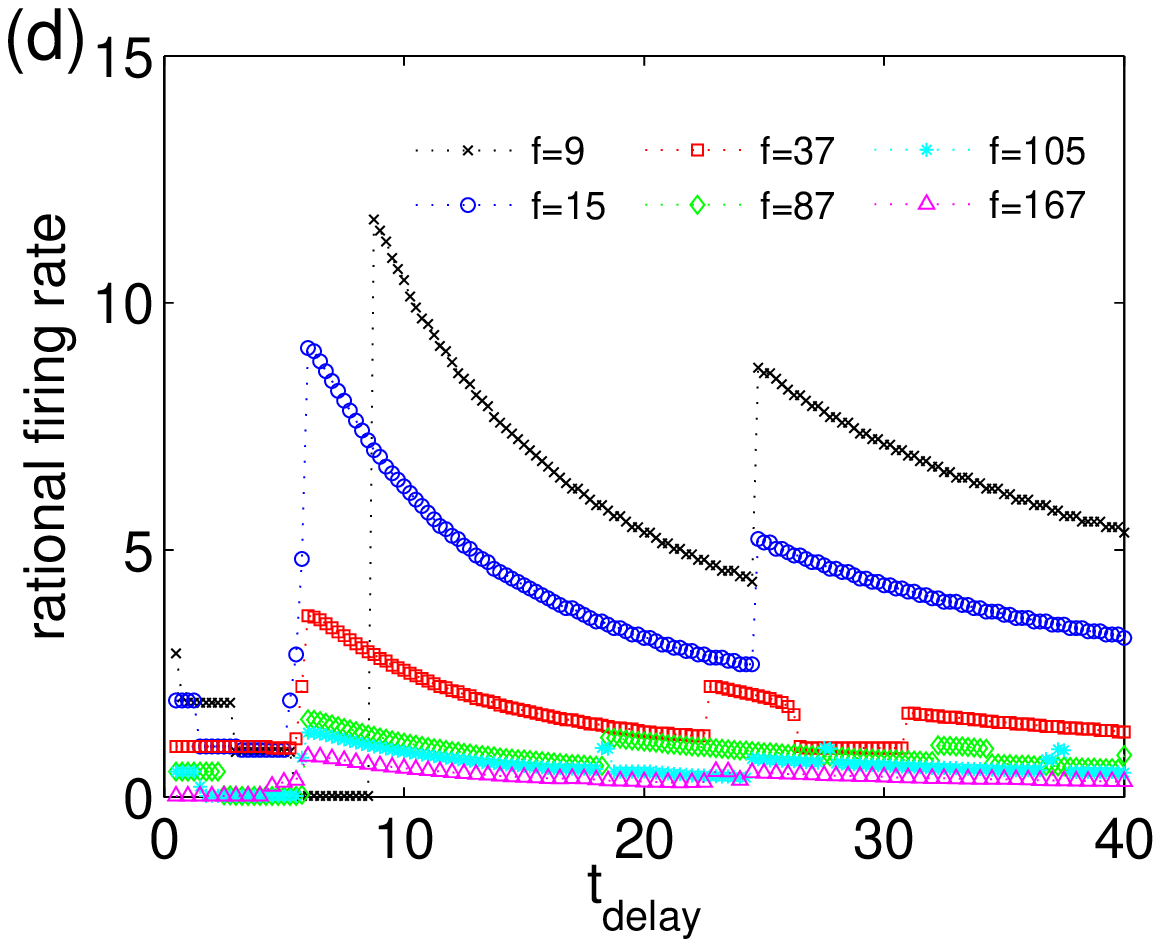}
\end{center}
\caption{Mode-locking of responses of an HH autaptic neuron with different delay times~\cite{wht_jtb}. The lines with different markers denote the firing rate of the neuron excited by different input frequencies. }
\label{jtb_fig1}
\end{figure}

Without an autapse, the mode-locking behaviors depend on the values of the stimulus frequencies and amplitudes. As shown in Fig.~1 in Ref~\cite{wht_jtb}, the Arnold tongues in the frequency and amplitude space show the overall features of various phase-locked states and provide the different $p:q$ (denoting output action potentials per input spikes) mode-locking regions. Ref~\cite{leesg} also shows the bifurcation mechanisms that create the boundaries of the complex mode-locking structure. 

The presence of an autapse substantially modifies the mode-locking patterns of the neuron. For the same sinusoidal stimulus, the neuron with an autapse can fire more or fewer action potentials than that of a neuron without an autapse. The activities of the neuron can also be driven into sub-threshold oscillation. The electrical autapse displays a regulating function that controls the mode-locking firing of the neuron. The time courses of the membrane potentials of the neuron with and without an autapse are shown in Fig.~2 of Ref.~\cite{wht_jtb}. With an autapse, the modification of the mode-locking firing pattern depends on the autaptic conductivity and the delay time. When the synaptic conductivity is small, the mode-locking $p:q$ is similar to that without an autaptic connection. When the synaptic conductivity is large, the neuron displays very complex mode-locking firing. The $p:q$ value of mode-locked firing increases from zero to a very large value when the frequency of the sinusoidal current increases from zero to the "threshold" frequency (the minimal input frequency for which the neuron fires an action potential given a fixed input amplitude). For further increases in the frequency of the sinusoidal current, the $p:q$ value of mode-locking firing decreases. In the case of extremely strong autaptic conductivity [see Fig.~3 (d) in Ref.~\cite{wht_jtb}], the autaptic activities completely disrupt the original mode-locking firing of the neuron without an autapse.

For continuous changes in the delay time, the model-locked firing of the neuron is also quite interesting. Fig.~\ref{jtb_fig1} shows the dependence of the mode-locking state on the delay time for different input frequencies and autaptic conductivities. For weak autaptic conductivities [Fig.~\ref{jtb_fig1}(a)], the $p:q$ values of mode-locking are similar to those without self-feedback for the short delay time. As the delay time increases, the values of $p:q$ mode-locking begin to fluctuate. The smaller the input frequency is, the larger the resulting fluctuation. For strong autaptic intensities [shown in Fig.~\ref{jtb_fig1}(b-d)], the $p:q$ values of mode-locking decrease with increasing delay time in a stepwise manner. With further increases in the delay time, the $p:q$ values of mode-locking suddenly jump to a very large value (larger than that without autaptic self-feedback) and then decrease smoothly. As the delay time increases, the $p:q$ value of mode-locking of a neuron with a high autaptic intensity exhibits periodic behavior.

With autaptic self-feedback, the responses of an HH neuron to a sinusoidal stimulus can have higher or lower $p:q$ mode-locking responses than that of a neuron without an autapse. These dynamical behaviors depend on the autaptic intensity and the delay time. Moreover, the presence of an autapse increases the range of values for which the HH neuron spiking is locked with the sinusoidal current. When the autaptic intensity is weak, the mode-locking behaviors show no marked changes. That is, an autapse with a weak intensity will have a negligible effect on the response of the neuron. For stronger autaptic intensities, the $p:q$ value of the mode-locking increases and then decreases as the delay time increases, exhibiting nearly periodic behaviors. Thus, for sufficiently strong autaptic intensity, changing the delay time provides better regulation of mode-locking than does changing the autaptic intensity. The autaptic connection may also provide a control option for adjusting mode-locked firing in a neural information process.

\section{Responses of a regular spiking neuron}
\subsection{Response to DC currents}

\begin{figure}
\begin{center}
\includegraphics[width=0.236\textwidth]{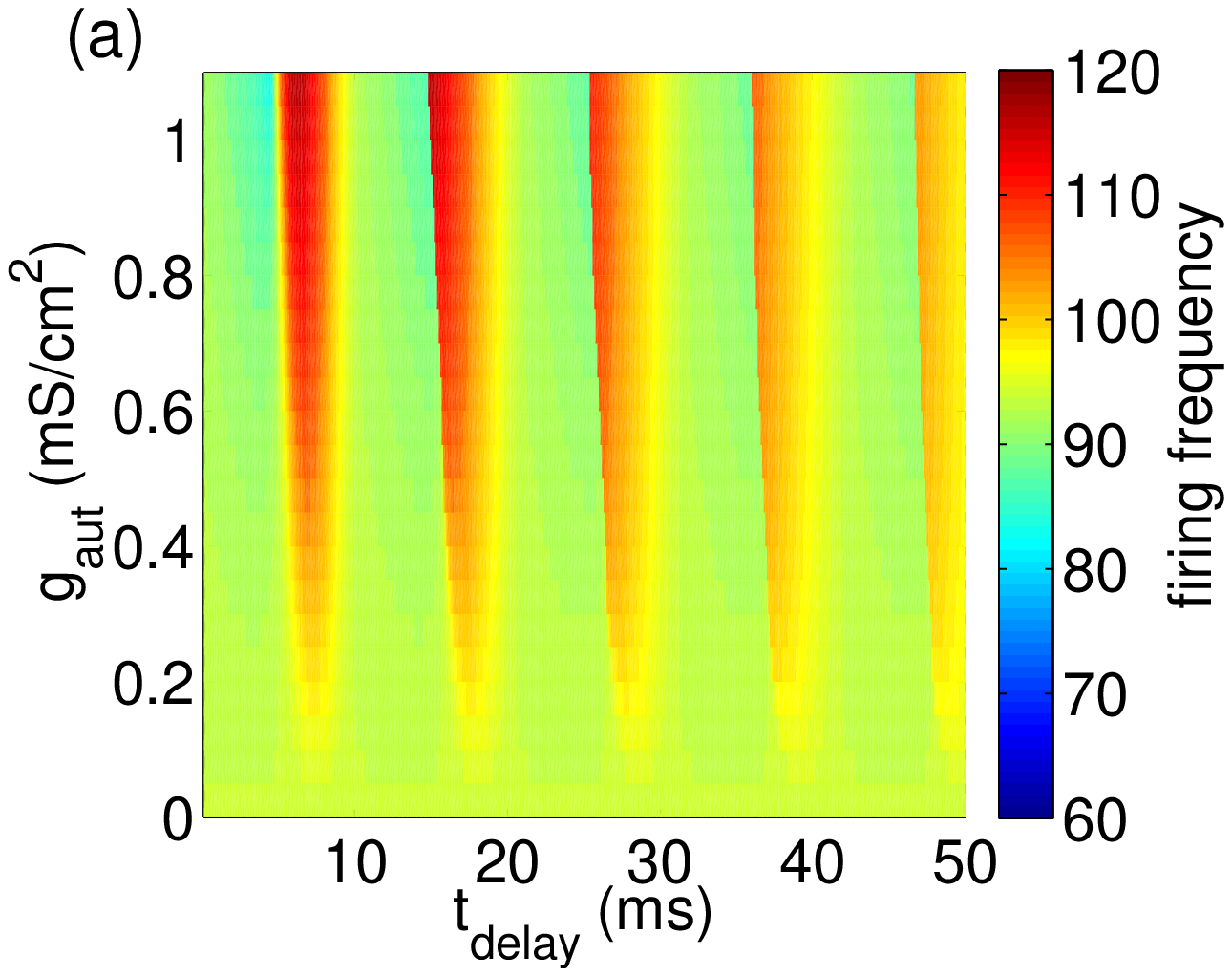}
\includegraphics[width=0.236\textwidth]{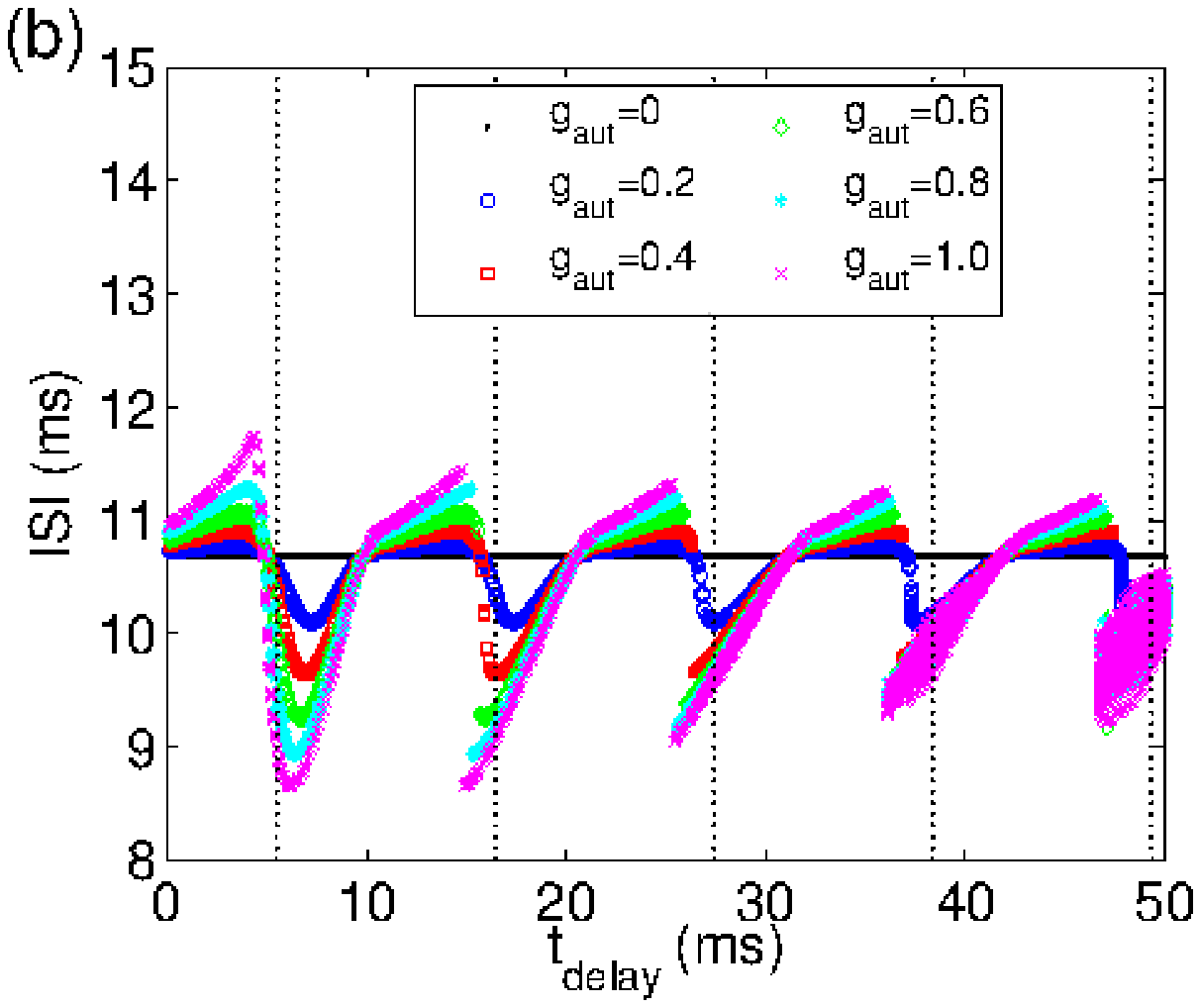}

\includegraphics[width=0.236\textwidth]{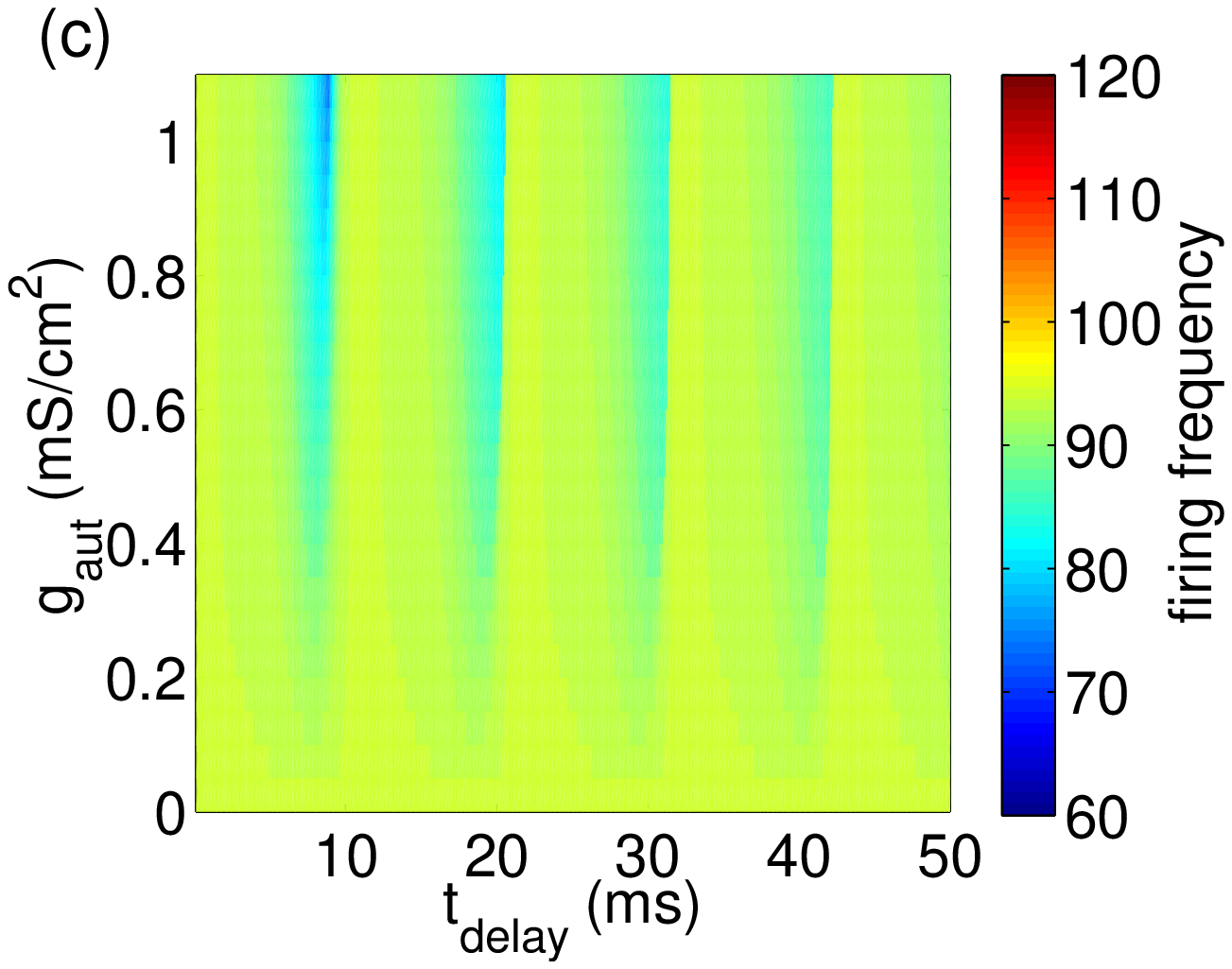}
\includegraphics[width=0.236\textwidth]{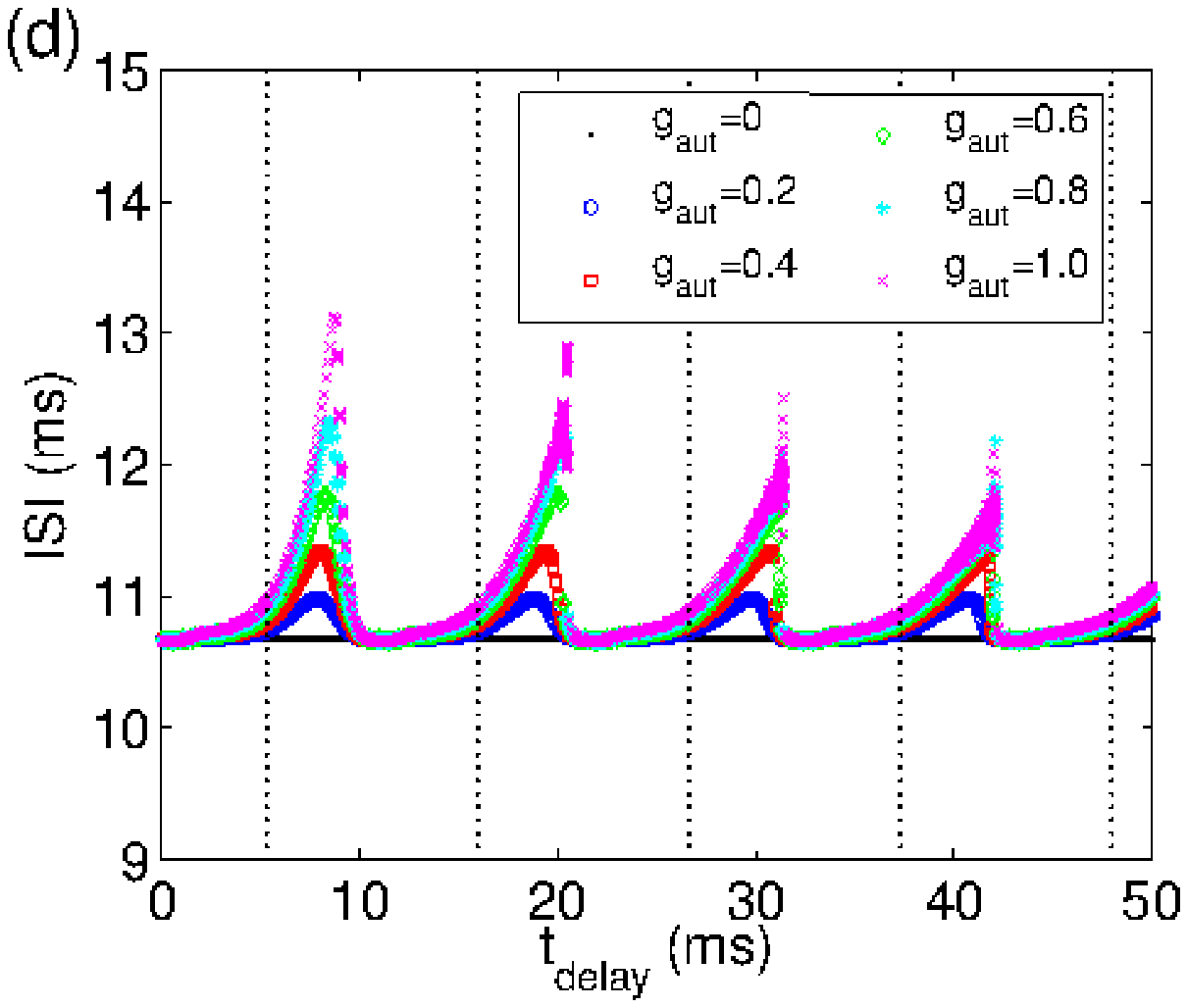}
\caption{(Color online) Firing frequencies ( (a),(c)) and ISI distribution ((b), (d)) of a neuron with a chemical autapse (upper panels: excitatory autapse; lower panels: inhibitory autapse) excited by a DC current~\cite{wht_chaos}. Dotted lines gives the odd multiples of half the intrinsic period of a single HH neuron excited by the corresponding DC current.} 
\label{dc5}
\end{center}
\end{figure}

When injected with a DC current, an HH neuron without an autapse displays regular spiking when the current intensity is larger than a critical current value. As for most neuronal models, increases in the current will increase the firing rate of the neuron, and the HH neuron undergoes a Hopf bifurcation. 
In the presence of an autapse, the output frequency can be higher or lower than the intrinsic frequency (the firing frequency of an HH neuron without an autapse) when the autaptic conductance and delay time are changed. Interestingly, the output frequency, as well as the output ISI distribution, shows periodic behaviors as the autaptic delay time increases. This periodicity of the changes in the output frequency (ISI distributions) in response to changes in the delay time is also similar to the intrinsic period of the neuron. When an HH neuron is connected to an electrical or excitatory chemical autapse, the firing rate and the ISI distribution of the response spike trains can both be either amplified or depressed. When the delay time approaches the odd multiples of the intrinsic half period of the firing of an HH neuron with an electrical or excitatory chemical autapse, the ISIs of the spike train are very different from the intrinsic ISIs. When the delay time approaches a multiple of the intrinsic period, the ISIs of the spike train are similar to the intrinsic ISIs. However, the inhibitory chemical autapse can only suppress the firing responses of the neuron. The ISIs of the response spike train of a neuron with an inhibitory autapse are always greater than the intrinsic period, whereas the corresponding response frequency is not less than the intrinsic response frequency. Current neurobiological experiments also reveal that autapses in the nervous system can be either self-inhibitory or self-excitatory, depending on their location on the neuron~\cite{Bekkers1998,bekkers2003,gulledge1,gulledge2}. 

\begin{figure}
\begin{center}
\includegraphics[width=0.224\textwidth]{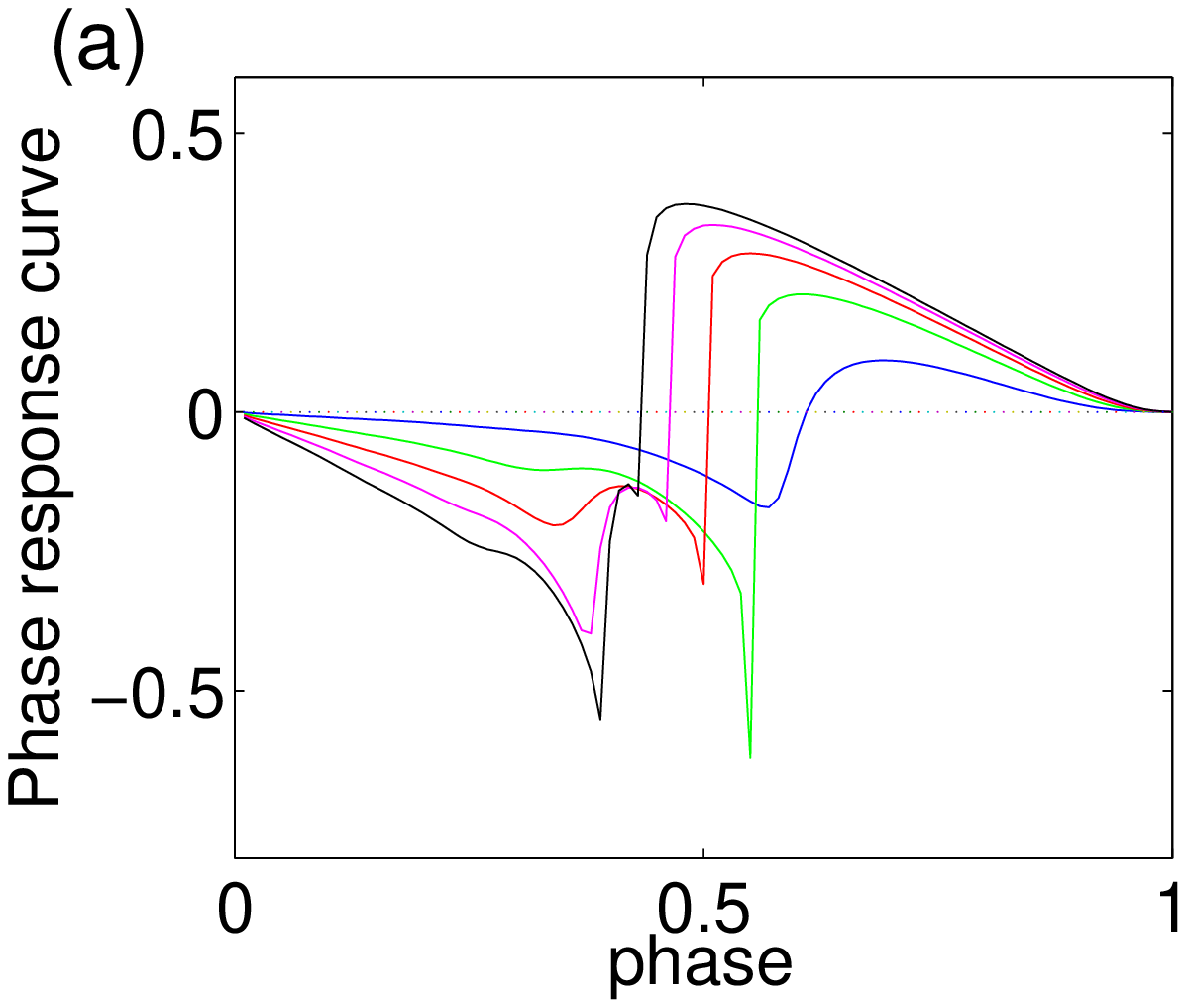}
\includegraphics[width=0.224\textwidth]{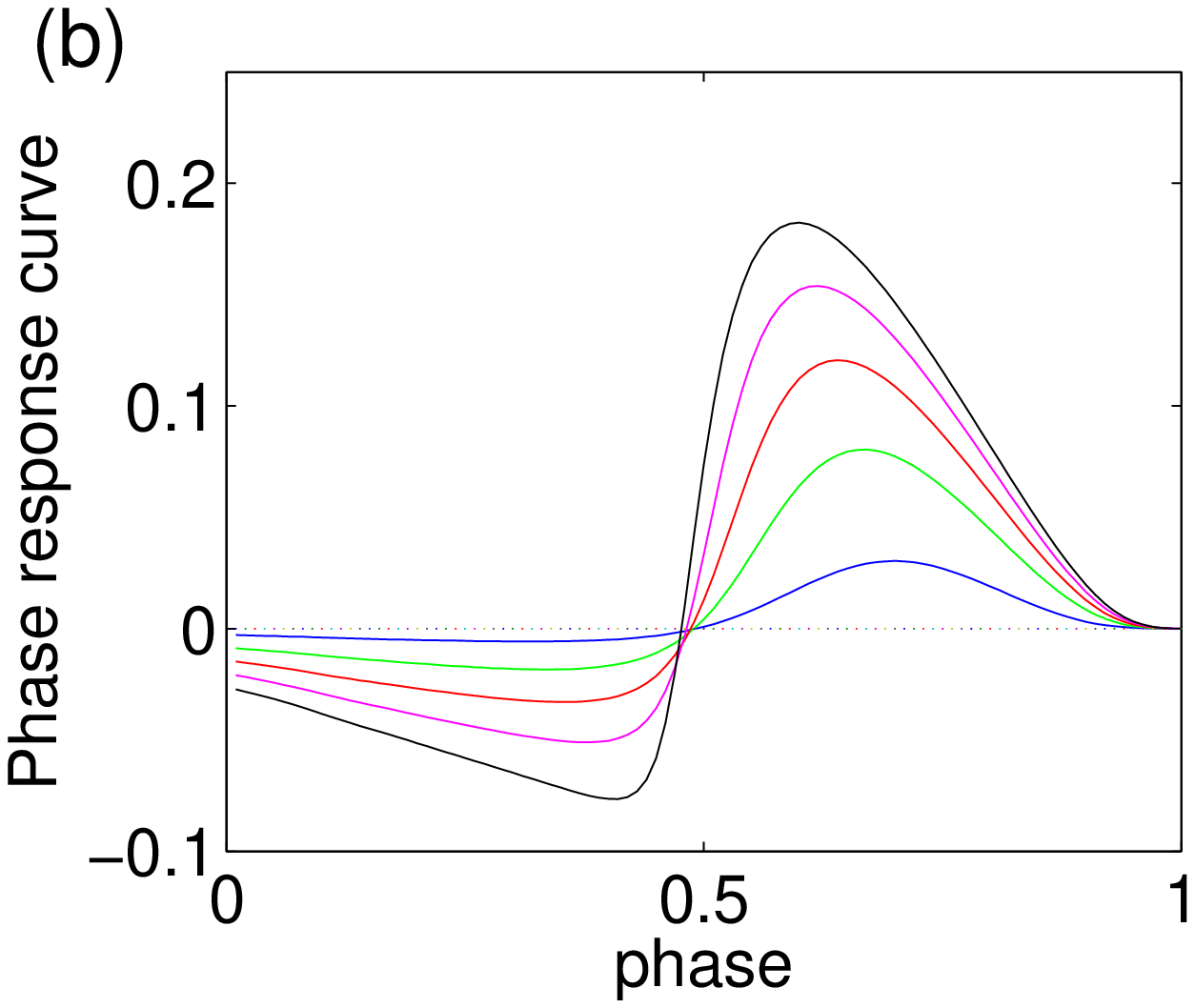}
\includegraphics[width=0.224\textwidth]{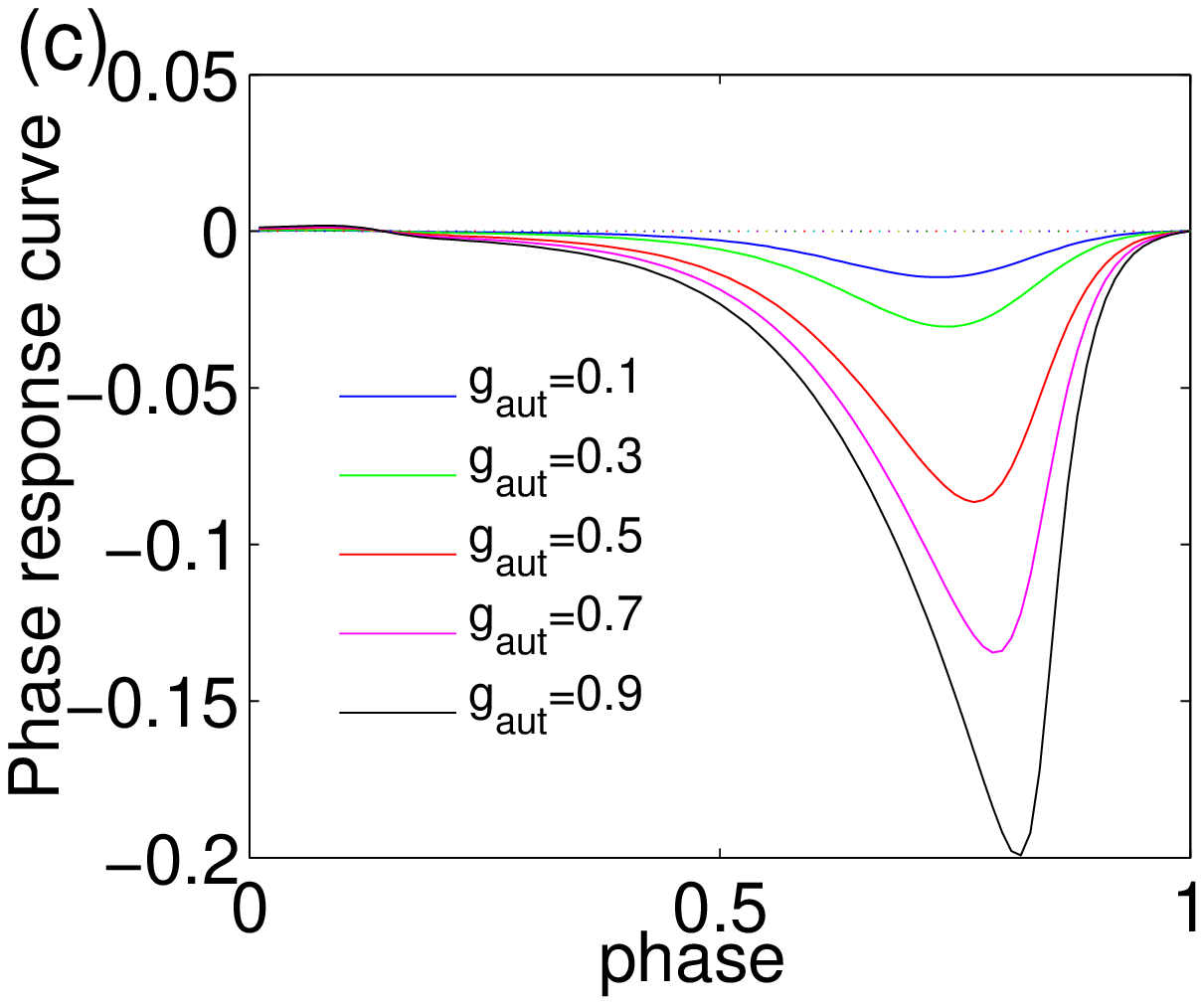}
\caption{(Color online) The phase-response curve of three types of autapses with different autaptic conductivity $g_{aut}$, (a) electrical autapse,  (b) excitatory chemical autapse, and (c) inhibitory chemical autapse. The external DC stimuli is set as 26 $\mu$A/cm$^2$.}
\label{rep_fig2}
\end{center}
\end{figure}

Because autaptic pulses perturb neural spiking through a process that is similar to that of an oscillating system with a fixed-delay feedback, the phase-response curve (PRC) theory can provide useful insight into the phenomena of the neuron-autapse system~\cite{glass1,glass2,glass3,glass4}. For an oscillating system, the PRC describes the phase shift that occurs in response to a brief external stimulus~\cite{Carmen1,Carmen2}. Fig.~\ref{rep_fig2} gives the phase-response curves of the three types of autapses. The PRCs of a neuron with an electrical autapse or an excitatory chemical autapse can be negative or positive, depending on the stimulus phase (delay time). Thus, an electrical autapse or excitatory chemical autapse could delay or advance the next spike. Therefore, the firing rate of a neuron with an electrical autapse or an excitatory chemical autapse can be higher or smaller than that of the neuron without an autapse. For the inhibitory chemical autapse, however, the PRC is always positive, because the delay time increased. That is, an inhibitory autapse always postpones the next spike. Thus, the firing rate of an HH neuron with an inhibitory autapse is not larger than that of the neuron without an autapse.

\begin{figure}
\begin{center}
\includegraphics[width=0.5\textwidth]{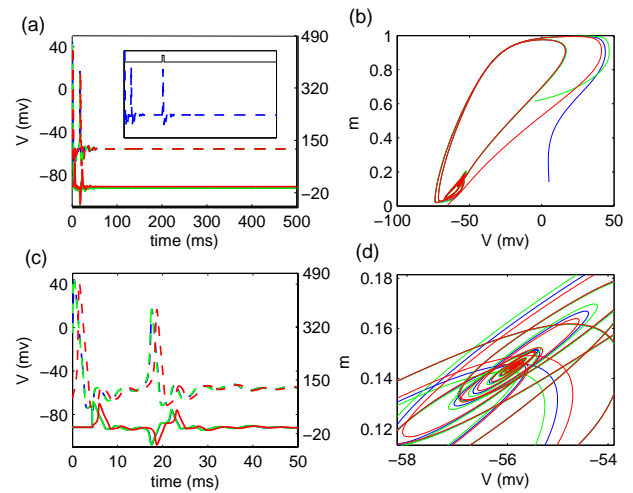}
\caption{(Color online) (a) Traces of the membrane potential (dashed lines) and autaptic current (solid lines)~\cite{wht_chaos}. (b) The phase portrait corresponds to (a). (c) and (d) are the enlarged plots of (a) and (b), respectively. The different colors indicate different integral initial conditions (The specific set of parameters can be found in Ref.~\cite{wht_chaos}). The inset in (a): The membrane potential of HH neuron-autapse system perturbed by a brief step pause (black solid line).} 
\label{rfig4_v}
\end{center}
\end{figure}

When the DC current increases, the HH neuron without an autapse undergoes a Hopf bifurcation from a quiescent state to a periodic spiking state. Although the limit cycle and the oscillation period of the neuron have been disturbed by the autaptic current, the neuron-autapse system still undergoes the Hopf bifurcation, generating a stable periodic orbit. Interestingly, an HH neuron with some special autaptic parameters does not fire regular action potentials and is attracted to the quiescent state. This result reveals the phenomenon of spiking death that is induced by an autaptic connection, which is more clearly shown in the membrane potential and autaptic current traces in Fig.~\ref{rfig4_v}. With an upper-threshold DC current, the neuron fires action potentials at first. After the delay time, the subsequent autaptic pulse reaches the neuron and drives the neuron to a fixed point. Moreover, the spiking death induced by the autapse is independent of the initial conditions of the neuron. As shown in the inset of Fig.~\ref{rfig4_v} (a), the neuron-autapse system can return to the fixed point even when the input is a brief external perturbation. The figure also shows that the spiking death induced by the autapse is stable. The Ref.~\cite{sjw} also shows a similar phenomenon: an HH neuron transitions from a limit cycle to a fixed point when the neuron is perturbed by an excitatory chemical synaptic pulse.

\subsection{Response to a random synaptic pulse-like input}

The assumed $\alpha$ form of the postsynaptic current model that is used in many works is perfect for generating pulse-like currents that are similar to the synaptic pulses observed experimentally and also enables easy modification of the ISI of the input current to investigate the input-output properties conveniently~\cite{wanght}. Injecting this type of synaptic pulse-like signal with different ISIs into the HH neuron, Hideo et al. have reported that the mean firing frequency decreases as the mean input ISI increases~\cite{Hideo,Borkowski}. When a neuron is injected with a synaptic-like pulsed current with random ISIs, the neuron displays interesting autapse-induced response behaviors~\cite{wht_chaos}. As the delay time increases, the frequency of the neuron with a sufficient electrical or excitatory chemical autapse shows nearly periodic behaviors. Such periodicity is not changed when the mean ISIs of the input synaptic current changes. When given the input synaptic pulses with a large mean value of ISIs, however, the response frequency of a neuron with an inhibitory autapse does not show periodic behavior when the autaptic delay time increases. 

\begin{figure}
\begin{center}
\includegraphics[width=0.45\textwidth]{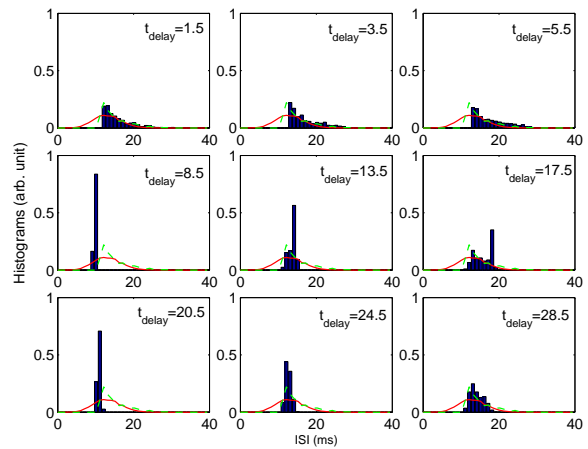}
\caption{(Color online) Histograms of output (blue bars) ISIs of a neuron with an electrical autapse~\cite{wht_chaos}. The green lines gives the output ISI of a neuron without an autapse and the red lines give the ISI of the random input pulse train.} 
\label{returnmap}
\end{center}
\end{figure}

Autaptic activities also influence the detailed response of the ISI distribution. For a short autaptic delay time, the output ISIs are distributed in an area [blue bars in Fig.~\ref{returnmap}] that is smaller than the input area (red lines in Fig.~\ref{returnmap}). When the delay time increases, the region in which the output ISIs are distributed does not change much compared with that for the neuron without an autapse (green lines in Fig.~\ref{returnmap}). For further increases in the delay time, the distribution of the output ISIs suddenly decreases almost to a single point ($t_{delay}=8.5$ ms in Fig.~\ref{returnmap}). For these autaptic parameters, the responding spike train is strongly regulated.
Then, the size of the distribution of ISIs increases slowly as the delay time is increased further.

When the size of the distribution of the output ISIs increases to its maximum (this maximum distribution area is smaller than the area for the low delay time condition $t_{delay}<5.5 \mathrm{ms}$), the distribution suddenly shrinks nearly to a point again (see $t_{delay}=20.5$ ms in Fig.~\ref{returnmap}). For the entire range of delay times, the distribution of the output ISIs periodically exhibits the above behaviors when the delay time increases. Thus, the delayed feedback activities of an autapse can act as a regulator that adjusts the ISIs of the output spike train. For some specific autaptic parameters, the resulting spike train of a neuron can be modified to obtain an almost regular spiking, even for a very random ISI input.

Without an autapse, the neuron will filter the spikes with a short ISI and thus show low-pass filtering behavior when the neuron is injected with a random synaptic pulse-like current~\cite{Hideo}. For a neuron with an autapse, long ISI pulses can be filtered in addition to the short ISI spikes. Thus, the neuron-autapse system displays complex filtering behaviors, including low-pass filtering and band-filtering behaviors.

When the delay time is short, an HH neuron with an electrical or excitatory chemical autapse acts as a low-pass filter similar to a neuron without an autapse and removes the spikes with short ISIs. When the autaptic delay time is long enough, the neuron displays band-pass filtering behavior and removes both the short- and long-ISI spikes. The cut-off value for the ISIs (either the minimum or the maximum ISIs that will be retained in the output spike train) can be changed by changing the autaptic parameters. For some specific autaptic parameters, the neuron can filter most of the input pulse, and the output spike train can be altered into a nearly regular spike train, even for an input with highly random ISIs. More interestingly, a neuron with an inhibitory chemical autapse can only act as a low-pass filter. Thus, the inhibitory chemical autapse does not have a significant effect on the frequency of the output spike train.

It can be conjectured that the filtering properties depend on the intrinsic properties of both the neuron and the autapse. For electrical or excitatory chemical autapses, the delay time will act as a border or as a cut-off value for the ISI filtering of synaptic pulses with a long ISI. The intrinsic filtering property of neurons removes the short input ISI pulses, and the time delay of the electrical or excitatory chemical autapse removes the long input ISI pulses. However, an inhibitory autapse delays the spikes of a neuron and thus filters the spikes that have short ISI pulses. Thus, the firing frequency of a neuron with an inhibitory autapse is not higher than that of a neuron without an autapse.

\section{Effect of noise on the firing dynamics of a neuron}

Neuronal noise is random electrical fluctuations within neuronal networks and affects the patterns of neural activity in a determinant way~\cite{white,Jacobson,wanghuiqiao,Destexhe}. In the presence of noise, autapses can also substantially affect the response of a neuron. In this section, we review previous studies on the interplay of autapses and noise on the firing of an autaptic neuron.

Considering the subthreshold dynamics of a neuron with interaction between autaptic-delayed feedback and noise, Masoller et al. investigated the firing patterns of an HH model of a thermoreceptor neuron with an electrical autaptic feedback in the presence of a Gaussian white noise~\cite{masoller}. In their studies, the neuron displays only subthreshold oscillations in the absence of feedback and noise. Their results show that the interaction among weak autaptic feedback, noise, and the subthreshold intrinsic activity is nontrivial. The subthreshold oscillation amplitude can be enhanced by the autapse in the presence of external noise, and this enhancement is more pronounced for certain delay values. For negative autaptic delay feedback, the firing rate can be lower than that of the noise-free situation, depending on the delay. This is because noise inhibits feedback-induced spikes by driving the neuronal oscillations away from the firing threshold. For positive autaptic delay feedback, there are regions of delay values where the noise-induced spikes are inhibited by the feedback; in this case, the autaptic feedback drives the neuronal oscillations away from the threshold.  

Li and his co-workers analyzed the effects of electrical autapses on the spiking dynamics of a stochastic HH neuron~\cite{liyy}, considering the stochastic gating of ion channels or the so-called intrinsic channel noise is considered. They found that the delayed feedback manifests itself in the occurrence of bursting and a rich multimodal interspike interval distribution, exhibiting a delay-induced reduction in the spontaneous spiking activity at characteristic frequencies. For small numbers of ion channels, the channel noise is sizable and the excitatory dynamics remain practically unaffected by the delay. However, smaller noise levels and stronger autaptic intensities induce different synchronization phenomena between the delay time and the intrinsic time scales. The delay time and the intrinsic time scales determine the number of spikes that will be induced and become subsequently locked during one delay epoch. 

Recently, Yilmaz and Ozer showed that the electrical autaptic delayed feedback either enhances or suppresses the weak signal detection, depending on the parameters of autapse and channel noise~\cite{ozer}. When the delay time is close to integer multiples of the period of the intrinsic oscillations, the autapse enhances the weak periodic signal detection for the optimal values of the intrinsic noise and the autaptic intensity. The system response also exhibits stochastic resonance behavior, depending on the autaptic intensity. Moreover, the weak signal detection capability of the HH neuron is  strongly dependent on the cell size and autaptic strength.

\section{Conclusion and outlook}
Self-delayed feedback significantly affects nonlinear dynamic systems because such self-feedback loops introduce a new time scale into the dynamics of the system~\cite{soriano,park,soriano}. The system of a neuron with an autapse contains two time scales: the intrinsic time scale of the neuron and the time scale of the autapse. However, the situation is more complex in the nervous system, because there are many more time scales, including the time scales of the neurons, the synapse, the environment, and the autaptic connections. This complexity raises the question of how the activity time of an autapse influences the activities of coupled neurons in the nervous system. 

Autapses provide self-feedback circuits that are common in the nervous system. Since the naming of autapses, many experimental studies have revealed that autapses play important roles in brain function~\cite{Bekkers1998,ikeda,bacci,Kimura,salinas1,ychen2008,quiroga}. The current theoretical studies on the topic of autaptic connections also reported the importance of autapses and the many novel dynamic behaviors induced by the autapse~\cite{rusin,liyy,hashemi,wht2014a,wht_jtb,wht_chaos}. Autapses provide a control option that can sufficiently adjust the firing behaviors of a neuron for any form of input stimulus, regardless of the neuron type.

Autapses offer a new mechanism for switching between quiescent, periodic and chaotic firing patterns in bursting neurons~\cite{wht2014a}. Additionally, the nervous system responds rapidly to an external stimulus based on the transition between neuron firing patterns~\cite{Belykh,erichsen}. Thus, the results of the firing pattern transition induced by an autapse also indicate that an autapse could act as an efficient tool for controlling the transition among different relevant neuronal activities in the nervous system.

For an HH neuron with an autapse, the firing frequency and interspike interval distributions of the output spike train show periodic behavior when the delay time is increased~\cite{wht_chaos}. When specific autaptic parameters are chosen, the response spike trains are nearly regular, and the ISI distribution covers a small area, even with a highly random input. This phenomenon emphassises the autapse-induced filtering behaviors of the neuron. These results about the autapse are useful for studying the control of a nonlinear system.

In neuronal systems, an autapse can take the form of recurrent excitation, i.e., the discharge of a neuron, possibly after passing through the axon, can subsequently induce an excitatory response in the same neuron. This self-excitation mechanism is important for maintaining persistent activity, particularly the feeding behavior in sea slugs. Leonel et al. also reported that such processes play functional roles in amplifying activity in neuronal assemblies, causing reverberating activity, inducing some form of memory, or generating rhythmic patterns, such as those in central pattern generators~\cite{gomez}. The previous experimental data also indicate that the brain tissue expresses a novel form of self inhibition, namely autaptic inhibitory transmission in Fast-Spike (FS) cells slow self activities in interneurons~\cite{Pawelzik}. Ma et al. also investigated the activities of a two-dimensional neural network containing autapses with different time delays and found that the autapses induced many novel collective behaviors~\cite{majun2}. The effect of autapse on the synchronization of neural network have also been conducted in a group of HH neurons with small world network structure~\cite{wu}. It was found that the neurons exhibit synchronization transitions as autaptic delay feedback is varied, and fine synchronized network activities occur when an optimal autaptic strength is chosen.  
Alberto Bacci and his colleagues studied the autaptic self-inhibition of basket cells and the role of the autaptic connections of disinhibition within cortical circuits and argued that autaptic feedback could have a dual function in temporally coordinating parvalbumin basket cells during cortical network activity~\cite{deleuze}. Another study also indicated that fast spiking neurons with autapses showed the strongest asynchronous release in brain slices obtained from patients with intractable epilepsy, and that such discharges may be involved in generating and regulating network activities, including epileptic activity~\cite{jang}.

Although there are many studies on the topic of the autapse, the precise function of autapses and their contribution to information processing are remain unclear. There are also many open questions both in the experimental and theoretical areas: 1) What is the role of autapses in coordinating network activities? i.e., FS cell autapses in adjusting fast network synchrony. 2) What are the molecular mechanisms underlying autaptic asynchronous release and what is its relevance during physiological and pathological network activities? 3) What are the functions of autapses in the information propagation in the neural circuit and neural network? Addressing these questions will also facilitate our understanding of the fundamental mechanisms governing several core functions of cortical activities.

\end{CJK}

\begin{thebibliography}{999}
\bibitem{kandel} Kandel E R, Schwartz J H, Jessell T M, Siegelbaum S A and Hudspeth A J 2000 \textit{Principles of Neural Science} (New York: McGraw-Hill Medical)

\bibitem{Bartos} Bartos M, Vida I and Jonas P 2007 \textit{Nat. Rev. Neurosci.} \textbf{8} 45

\bibitem{Bennett} Bennett M V L and Zukin R S 2004 \textit{Neuron} \textbf{41} 495

\bibitem{Connors} Connors B W and Long M A 2004 \textit{Annu. Rev. Neurosci.} \textbf{27} 393

\bibitem{vander}Van der Loos H and Glaser E M 1972 \textit{Brain Res.} \textbf{48} 355

\bibitem{Bekkers1998} Bekkers J M 1998 \textit{Curr. Biol.} \textbf{8} R52

\bibitem{yamaguichi} Yamaguchi K 2008 \textit{Autapse Encyclopedia of Neuroscience} (M. D. Binder ed.), N Hirokawa and U Windhorst (Springer Berlin Heidelberg) pp 229-32

\bibitem{lubke} L\"{u}bke J, Markram H, Frotscher M and Sakmann B 1996 \textit{J. Neurosci. } \textbf{16} 3209

\bibitem{flight} Flight M H 2009 \textit{Nat. Rev. Neurosci. } \textbf{10} 316 

\bibitem{branco} Branco T and Staras K 2009 \textit{Nat. Rev. Neurosci.} \textbf{10}  373 

\bibitem{Kimura} Kimura F, Otsu Y and Tsumoto T 1997 \textit{J. Neurophysiol.} \textbf{77} 2805 

\bibitem{tamas} Tam\'{a}s G, Buhl E H and Somogyi P 1997  \textit{J. Neurosci.} \textbf{17} 6352

\bibitem{saada} Saada R, Miller N, Hurwitz I and Susswein A J 2009 \textit{Curr. Biol. } \textbf{19}  479 

\bibitem{ikeda} Ikeda K and Bekkers J M 2006 \textit{Curr. Biol. } \textbf{16} R308

\bibitem{bacci} Bacci A and Huguenard J R 2006 \textit{Neuron} \textbf{49} 119

\bibitem{bekkers09} Bekkers J M 2009 \textit{Curr. Biol.} \textbf{19}, R296

\bibitem{salinas1} Salinas E and Sejnowski T J 2001 \textit{Nat. Rev. Neurosci.} \textbf{2} 539

\bibitem{ychen2008}Chen Y, Yu L and Qin S M 2008 \textit{Phys. Rev. E} \textbf{78} 051909

\bibitem{chenyl} Chen Y, Zhang H, Wang H, Yu L and Chen Y 2013 \textit{PLoS ONE} \textbf{8} e56822

\bibitem{quiroga} Quiroga R  Q  and Panzeri S 2009 \textit{Nat. Rev. Neurosci.} \textbf{10} 173

\bibitem{davidson} Davidson E. and Levin M. 2005 \textit{Proc. Natl. Acad. Sci. U.S.A.} \textbf{102} 4935

\bibitem{zhanghui} Zhang H, Chen Y and Chen Y 2012 \textit{PLoS ONE} \textbf{7} e51840

\bibitem{cabrera} Cabrera J L and Milton J G 2002 \textit{Phys. Rev. Lett.} \textbf{89} 158702

\bibitem{bonan} Bonan G B 2008 \textit{Science} \textbf{320} 1444

\bibitem{post} Postlethwaite C M and Silber M 2007 \textit{Phys. Rev. E} \textbf{76} 056214

\bibitem{sethia} Sethia G C, Kurths J, and Sen A 2007 \textit{Phys. Lett. A} \textbf{364} 227

\bibitem{waik} Chen W-K 2005 \textit{Circuit Analysis and Feedback Amplifier Theory} (Bocan Raton, FL: CRC Press)

\bibitem{Gaudreault} Gaudreault M, Drolet F and Viñals J 2012 \textit{Phys. Rev. E} \textbf{85} 056214

\bibitem{Ahlborn} Ahlborn A and Parlitz U 2004 \textit{Phys. Rev. Lett.} \textbf{93} 264101

\bibitem{Balanov} Balanov A G, Janson N B and Sch\"{o}ll E 2005 \textit{Phys. Rev. E} \textbf{71} 016222

\bibitem{popovych} Popovych O V, Hauptmann C and Tass P A 2006 \textit{Biol. Cybern.} \textbf{95} 69


\bibitem{rusin} Rusin C G, Johnson S E, Kapur J and Hudson J L 2011 \textit{Phys. Rev. E} \textbf{84} 066202


\bibitem{prager} Prager T, Lerch H P, Schimansky-Geier L and Sch\"{o}ll E 2007 \textit{J. Phys. A} \textbf{40} 11045


\bibitem{liyy}  Li Y, Schmid G, Hänggi P and Schimansky-Geier L 2010 \textit{Phys. Rev. E} \textbf{82} 061907

\bibitem{hashemi}  Hashemi M, Valizadeh A and Azizi Y 2012 \textit{Phys. Rev. E} \textbf{85} 021917


\bibitem{wht2014a} Wang H, Ma J, Chen Y and Chen Y 2014 \textit{Commun. Nonlinear Sci. Numer. Simul.} \textbf{19} 3242

\bibitem{schacter} Schacter D L, Gilbert D T, Wegner D M and Nock M K 2014 \textit{Psychology} (New York, NY: Worth Publishers)

\bibitem{jang}  Jiang M, Zhu J, Liu Y, Yang M, Tian C, Jiang S, Wang Y, Guo H, Wang K and Shu Y 2012 \textit{PLoS Biol.} \textbf{10} e1001324

\bibitem{majun1} Jun Ma H Q 2015 \textit{Int. J. of Mod. Phys. B} \textbf{29} 1450239

\bibitem{Belykh} Belykh I, de Lange E and Hasler M 2005 \textit{Phys. Rev. Lett.} \textbf{94} 188101

\bibitem{buric} Buri\'{c} N, Todorovi\'{c} K and Vasovi\'{c} N 2008 \textit{Phys. Rev. E} \textbf{78} 036211 

\bibitem{wht_jtb} Wang H, Sun Y, Li Y and Chen Y 2014 \textit{J. Theor. Biol.} \textbf{358} 25

\bibitem{wht_chaos} Wang H, Wang L, Chen Y and Chen Y 2014 \textit{Chaos} \textbf{24} 033122

\bibitem{william}Connelly W M and Lees G 2010 \textit{J. Physiol.} \textbf{588}, 2047

\bibitem{cocatre}Cocatre-Zilgien JH, Delcomyn F 1992 \textit{J. Neurosci. Meth.} \textbf{41} 19 

\bibitem{mainen}  Mainen Z F and Sejnowski T J 1996 \textit{Nature} \textbf{382} 363

\bibitem{izikevich2000}  Izhikevich E M 2000 \textit{Int. J. Bifurcat. Chaos} \textbf{10} 1171

\bibitem{kepecs2002}   Kepecs A, Wang X-J and Lisman J 2002 \textit{J. Neurosci.} \textbf{22} 9053

\bibitem{guhuaguang} Gu Hua-Guang,Zhu Zhou,Jia Bing. 2011 \textit{Acta Physica Sinica} \textbf{60} 100505

\bibitem{yuht} Yu H-T, Wang J, Deng B and Wei X-L 2013 \textit{Chin. Phys. B} \textbf{22} 018701

\bibitem{wagenaar}   Wagenaar D A, Pine J and Potter S M 2006 \textit{ BMC Neurosci.} \textbf{7} 11

\bibitem{wangj} Yu H, Wang J, Deng B, Wei X, Wong Y K, Chan W L, Tsang K M, Ziqi Y. et al. 2011 \textit{Chaos} \textbf{21} 013127

\bibitem{tang}   Tang G, Xu K and Jiang L 2011 \textit{Phys. Rev. E} \textbf{84} 046207

\bibitem{wangxj}   Wang X-J 1993 \textit{Physica D} \textbf{62} 263

\bibitem{gnzalez}   Jm G-M 2003 \textit{Chaos} \textbf{13} 845

\bibitem{innocenti}   Innocenti G, Morelli A, Genesio R and Torcini A 2007 \textit{Chaos} \textbf{17} 043128

\bibitem{innocenti2009}   Innocenti G and Genesio R 2009 \textit{Chaos} \textbf{19} 023124

\bibitem{leesg}  Lee S-G and Kim S 2006 \textit{Phys. Rev. E} \textbf{73} 041924

\bibitem{chey}   Che Y-Q, Wang J, Si W-J and Fei X-Y 2009 \textit{Chaos, Solitons} \& \textit{Fractals} \textbf{39} 454

\bibitem{wanght} Wang H, Wang L, Yu L and Chen Y 2011 \textit{Phys. Rev. E.} \textbf{83} 021915

\bibitem{fellous} Fellous J-M, Houweling, A R, Modi, R H, Rao, R P N, Tiesinga P H E and Sejnowski T J2001 \textit{J Neurophysiol.} \textbf{85} 1782 

\bibitem{hodgkin} Hodgkin A L and Huxley A F 1952 \textit{The J. of Physiol. } \textbf{117} 500


\bibitem{bekkers2003}Bekkers J M 2003  \textit{Curr. Biol.} \textbf{13}, R433

\bibitem{gulledge1}Gulledge A T  and Stuart G J 2003 \textit{Neuron} \textbf{37}, 299

\bibitem{gulledge2}Saada-Madar R, Miller N and  Susswein A J 2012 \textit{J. Mol. Histol.} \textbf{43} 431

\bibitem{glass1} Lewis J, Bachoo M, Glass L, and Polosa C 1987 \textit{Phys. Lett. A } \textbf{125}, 119 
 
\bibitem{glass2} Glass L and Zeng W-Z 1990 \textit{Ann.NY Acad.Sci. } \textbf{591}  316

\bibitem{glass3} Lewis J E, Glass L, Bachoo M, and Polosa C 1992 \textit{J. Theor. Biol.} \textbf{159} 491

\bibitem{glass4} Kunysz A M, Shrier A, and Glass L 1997 \textit{Am. J. Physiol.} \textbf{273} C331


\bibitem{Carmen1}Smeal R M, Ermentrout G B, and White J A  2010 \textit{Trans. R. Soc. B} \textbf{365} 2407 

\bibitem{Carmen2}  Krogh-Madsen T, Butera R, Ermentrout G B, and Glass L, in Phase Response Curves in Neuroscience, edited by N.W. Schultheiss,  Prinz A A., and Butera R J (Springer New York, 2012), pp. 33

\bibitem{sjw} Wang S-J, Xu X-J, Wu Z-X, Huang Z-G, and Wang Y-H 2008 \textit{Phys. Rev. E} \textbf{78} 061906

\bibitem{Hideo} Hasegawa H 2000 \textit{Phys. Rev. E} \textbf{61} 718
\bibitem{Borkowski} Borkowski L S \textit{Phys. Rev. E} \textbf{80} 051914

\bibitem{white} White J A, Rubinstein J T and Kay A R 2000 \textit{Trends in Neurosci.} \textbf{23} 131

\bibitem{Jacobson} Jacobson G A, et al. 2005 \textit{J. Physiol.} \textbf{564} 145

\bibitem{wanghuiqiao} Wang H, Yu L and Chen Y 2009 \textit{Acta Phys. Sin.} \textbf{58} 5070

\bibitem{Destexhe} Destexhe A and Rudolph-Lilith M 2012 Neuronal Noise (New York: Springer)

\bibitem{masoller} Masoller C, Torrent M C and Garcia-Ojalvo J 2008 \textit{Phys. Rev. E} \textbf{78} 041907

\bibitem{ozer} Yilmaz E and Ozer M 2015 \textit{Physica A} \textbf{421} 455

\bibitem{soriano} Soriano M C, Garcia-Ojalvo J, Mirasso C R and Fischer I 2013 \textit{Rev. Mod. Phys.} \textbf{85} 421

\bibitem{park} Park J-H, Huh S-H, Kim S-H, Seo S-J and Park G-T 2005 \textit{IEEE Trans. Neural. Netw.} \textbf{16} 414

\bibitem{erichsen} Erichsen R J, Brunnet L G. 2008  \textit{Phys. Rev. E} \textbf{78} 061917

\bibitem{gomez} G\'{o}mez L, Budelli R and Pakdaman K 2001 \textit{Phys. Rev. E} \textbf{64} 061910

\bibitem{Pawelzik} Pawelzik H, Hughes D I and Thomson A M 2003 \textit{J Physiol} \textbf{546} 701

\bibitem{majun2} Qin H, Wu Y, Wang C and Ma J 2015 \textit{Commun. Nonlinear. Sci. Numer. Simulat.} \textbf{23} 164

\bibitem{wu} Wu Y, Gong Y and Wang Q 2015 \textit{Chaos} \textbf{25} 043113

\bibitem{deleuze} Deleuze C, Pazienti A and Bacci A 2014 \textit{Curr. Opin. Neurobiol.} \textbf{26} 64

\end{thebibliography}
\end{document}